\documentclass[aps,prb,twocolumn,showpacs,superscriptaddress,groupedaddress]{revtex4}
\usepackage{graphicx}
\usepackage{fancyhdr}
\usepackage{amsmath}

\begin{document}

\title{Resonant state selection in synthetic ferrimagnets}

\author{B. C. Koop}
\affiliation{Royal Institute of Technology, 10691 Stockholm, Sweden}
\author{Yu. I. Dzhezherya} 
\author{K. Demishev}
\author{V. Yurchuk}
\affiliation{Institute of Magnetism, Ukrainian Academy of Sciences, Kiev, Ukraine}
\author{D. C. Worledge}
\affiliation{IBM T. J. Watson Research Center, Yorktown Heights, NY 10598, USA}
\author{V. Korenivski}
\altaffiliation{Corresponding author: vk@kth.se}
\affiliation{Royal Institute of Technology, 10691 Stockholm, Sweden}

\date{\today}

\begin{abstract}
Resonant activation of a synthetic antiferromagnet (SAF) is known to result in a dynamic running state, where the SAF's symmetric spin-flop pair continuously rotates between the two antiparallel ground states of the system, with the two magnetic moments in-phase in the so-called acoustical spin-resonance mode. The symmetry of an ideal SAF does not allow, however, to deterministically select a particular ground state using a resonant excitation. In this work, we study asymmetric SAF's, or synthetic ferrimagnets (SFi), in which the two magnetic particles are different in thickness or are biased asymmetrically with an external field. We show how the magnetic phase space of the system can be reversibly tuned, post-fabrication, between the antiferro- and ferri-magnetic behavior by exploiting these two asymmetry parameters and applying a uniform external field. We observe a splitting of the optical spin-resonance for the two ground states of the SFi system, with a frequency spacing that can be controlled by a quasistatic uniform external field. We demonstrate how the tunable magnetic asymmetry in SFi allows to deterministically select a particular ground state using the splitting of the optical spin-resonance. These results offer a new way of controlling the magnetic state of a spin-flop bilayer, currently used in such large scale applications as magnetic memory. 
\end{abstract}

\maketitle

\section{introduction}

\noindent Magnetic memory technology employs the effects giant magneto\-resis\-tance \cite{fert,grunberg} and tunneling magneto\-resis\-tance \cite{julliere,moodera} in spin-valves type sensors \cite{dieny}. In these, the switching of the magnetic state is achieved by using either external magnetic field \cite{gerrits,bergman,back,kaka} or spin-transfer-torque \cite{slonczewski,wu,myers,brataas} writing. In this work we demonstrate a reliable way to switch a magnetic nanodevice employing microwave excitation in resonance with the spin eigen-modes in the device, which act to amplify the action of the microwave field.

Synthetic antiferromagnets consist of two identical dipole-coupled magnetic layers separated by a thin spacer. The dipole coupling results in an antiparallel (AP) alignment of the magnetic moments in the ground state, which makes them similar to classical atomic antiferromagnets but with a tunable coupling strength between the magnetic moments set in fabrication by a suitable choice of the material parameters and geometry. When the lateral shape is elliptical, the system has two AP ground states, in which the moments are aligned along the easy-axis (EA) of the ellipse. Due to the symmetry, the energy of the two AP ground states is degenerate. 

The quasistatic \cite{stat_sw1,stat_sw2,stat_sw3} and dynamic \cite{res_sw1,res_sw2,res_sw3,res_sw4} behavior of SAFs has been studied extensively. Due to the nearly perfect flux closure in the AP state, SAFs are characterized by high stability against thermal agitation \cite{thermal}. It was recently shown that the spin dynamics contains collective acoustical and optical resonance modes \cite{kono}. It has also been shown that excitation at the optical resonance frequency can result in switching between the two AP ground states \cite{cherepov_prl}, typically resulting in multiple switching or a spin-running state, since the resonant switching in an ideal SAF is symmetric for the two AP ground states so a specific state cannot be selected deterministically.

In order to enable use of resonant excitation for selecting a specific ground state, which is highly desirable for applications, the spin-resonance spectrum can be tuned such that, for example, resonant switching occurs at different frequencies for the two ground states. For this the spin-flop bilayer must have suitable magnetic asymmetry, such as asymmetry in the the magnitude of the two magnetic moments or asymmetric field biasing of the layers externally so each layer experiences a different Zeeman field. The former can be achieved by making the two layers of the same material but slightly different thickness or of the same thickness but of materials with different saturation magnetization.  The latter can be built-in into the nominally flux-closed reference layer of the read out junction\cite{fringe}. These two magnetic asymmetry contributions can also be combined, which we do in this study for achieving greater flexibility in controlling the energetics and spin dynamics in the SFi system. 

In this paper we develop a method of deducing the individual magnetic asymmetry contributions in SFi, post-fabrication, and show how they can be used to reversibly tune the magnetic behavior of the system between SAF-like and SFi-like. We further show how an external uniform magnetic field can be used to control the symmetry of the magnetic double-well potential as well as the splitting of the optical spin resonance in SFi with a thickness imbalance and asymmetric local field-bias.

\begin{figure*}
\includegraphics[width=1\linewidth]{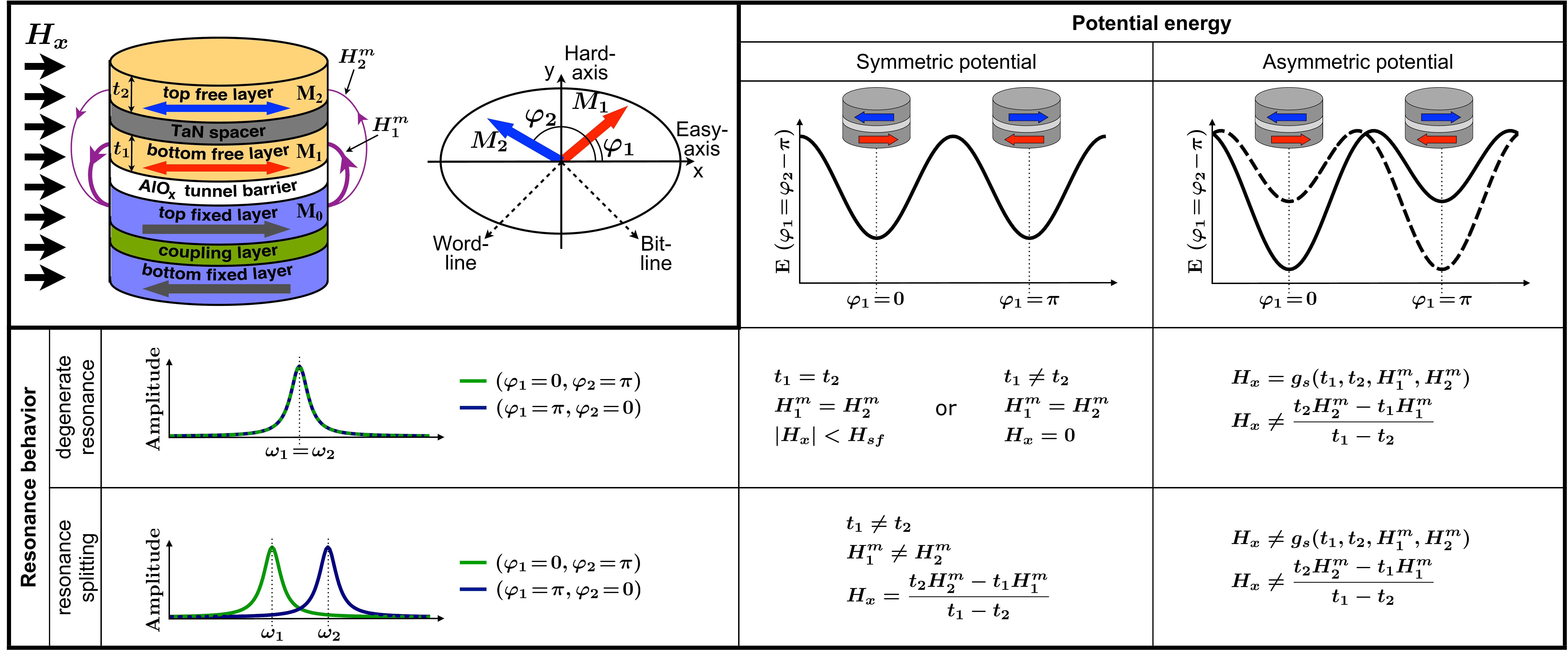}
\caption{(Color online) The top-left panel illustrates the layout of the nanopillar, which integrates an SFi bilayer and a nearly flux-closed read-out layer, separated by an Al-O tunnel junction, subject to a quasistatic magnetic field $H_x$ directed along the easy axis; the top-view schematic shows the magnetic layout of the bilayer, with the two dipole-coupled moments at angles $\varphi_1$ and $\varphi_2$. The optical spin resonance for the two ground states is illustrated in the bottom-left panel for a symmetric bilayer (SAF; single resonance peak) and an asymmetric bilayer (SFi; resonance doublet). The magnetic asymmetry is achieved by a thickness imbalance ($t_1 \neq t_2$) and/or asymmetric fringing field from the reference layer acting on the individual layers in the spin-flop bilayer ($H_1^m \neq H_2^m$). The top-right panels illustrate the symmetric and asymmetric magnetic potentials, generally corresponding to SAF and SFi. Interestingly, as shown in detail analytically in the text below, suitable magnetic asymmetry combined with a uniform external biasing field ($H_x$) can be used to reversibly tune the system between SAF-like (symmetric potential minima and a single resonance) and SFi-like (asymmetric potential and a resonance doublet) behavior. $H_x=H_{sf}$ is the spin-flop field of the bilayer, at which one of the minima vanishes completely and the two macrospins switch into a scissor state. The degeneracy in the magnetic potential and the sign and magnitude of the resonance splitting depend on the sign and magnitude of the external uniform biasing field (see text for details), which provides an easy and powerful tool for mapping out and controlling the static and dynamic properties of the system.}
\label{cartoon}
\end{figure*}

\section{SFi eigen-modes}

\noindent In this section we develop specific predictions for spin dynamics in SFi, which are then used to guide the experiment. In particular, we find analytically that the optical spin resonance in the system splits into a doublet, which allows controlled selection of the individual ground states using resonant microwave pulses modulated in frequency. We start by illustrating in Fig. \ref{cartoon} the energetics of SFi, including the key symmetry criteria for SAF-like versus SFi-like magnetic behavior.

In the macrospin approximation, appropriate for our sub-micrometer junctions, with the z-axis out-of-plane and the in-plane easy-axis (EA) due to shape anisotropy along $x$, the total magnetic energy$\;(E)$ of the bilayer can be written as \cite{stat_sw3}
\begin{align}
\frac{E}{2\pi M_S^2}=&\sum_{i,j=1;i\neq j}^2\!\!\!{V_i\left\{N_{ix}m_{ix}^2+(N_{iy}+h_{u})m_{iy}^2+N_{iz}m_{iz}^2\right.}\nonumber\\
&+\gamma_{jx}m_{1x}m_{2x}+\gamma_{jy} m_{1y}m_{2y}+\gamma_{jz} m_{1z}m_{2z}\label{E}\\
&{\left.-2(h_x+h_{ix}^m)m_{ix}-2h_y\cos{(\omega t)}m_{iy}\right\}},\nonumber
\end{align}

\noindent where $M_S$ is the saturation magnetization of the material (here taken to be same in both layers); $V_i=\pi abt_i/4$ the volume of layer $i$ with $a,\,b$ and $t_i$ the length, width, and thickness of that layer, respectively; $N_{i\alpha}$ the demagnetizing factor of layer $i$ in the $\alpha$-direction, with $\sum_\alpha N_{i\alpha}=1$ and $N_{i\{x,y\}}=n_{\{x,y\}}t_i/b$, and $n_{\{x,y\}}$ the reduced demagnetizing factors \cite{worledge2}; $\textbf{m}_i$ the cartesian unit magnetization vector of the $i$-th layer; $h_u=H_u/4\pi M_S$ the normalized intrinsic uniaxial anisotropy field along the easy-axis; $h_y=H_{y}/4\pi M_S$ the normalized amplitude of the applied microwave field in the y-direction; $\omega$ the frequency of the microwave field; $h_x=H_{x}/4\pi M_S$ the normalized external quasistatic easy-axis field; $h_i^m=H_i^m/4\pi M_S$ the normalized asymmetric biasing field (fringing field) acting on layer $i$ in the $x$-direction; $\gamma_{i\alpha}=r_\alpha N_{i\alpha}$ the inter-layer exchange field from layer $i$ acting on the other layer in the $\alpha$-direction \cite{dipole_couple}. In our case $\gamma_{i\alpha}$ is solely of dipolar nature. For simplifying the derivation to follow it is convenient to rewrite all terms proportional to $t_i$ through 
\begin{equation}
Q_i=\left[1-(-1)^i\varepsilon\right]Q,\quad \varepsilon=(t_1-t_2)/(t_1+t_2),\nonumber
\end{equation}
such that $Q_i=N_{ix},N_{iy},\gamma_{ix},\gamma_{iy}, V_i$.

The natural coordinate system for SFi is polar, in which
\begin{equation}
\textbf{m}_i=(\,\cos{\varphi_i}\,\sin{\theta_i},\quad\sin{\varphi_i}\,\sin{\theta_i},\quad\cos{\theta_i}\,).\label{polar}\nonumber
\end{equation}
For thin magnetic particles elongated in-plane in $x$, the following holds:

\begin{equation}
N_{ix}<N_{iy}\ll N_{iz}\approx 1;\quad\theta_i=\pi/2+\xi_i,\:\: |\xi_i|\ll1,\label{theta_i}\nonumber
\end{equation}
and $\cos(\theta_i)\approx-\xi_i$. Quantities $r_z$, $\xi_1$, $\xi_2$ are small parameters in the problem, therefore $\gamma_z\xi_i\xi_j$ can be neglected:
\begin{widetext}
\vspace{-10pt}
\begin{equation}
\begin{aligned}
\frac{W-W_0}{\pi M_S^2}=&-(1+\varepsilon^2)\left(N_y-N_x+\frac{h_u}{1+\varepsilon^2}\right)\cos{2\Phi}\cos{2\chi}+2\varepsilon\left(N_y-N_x+\frac{h_u}{2}\right)\sin{2\Phi}\sin{2\chi}\nonumber\\
&\quad-(1-\varepsilon^2)\left[(\gamma_y-\gamma_x)\cos{2\Phi}+(\gamma_y+\gamma_x)\cos{2\Phi}\right]-4(h_x+h_f+\varepsilon h_d)\cos{\Phi}\cos{\chi}\\
&\quad+4[\varepsilon(h_x+h_f)+h_d]\sin{\Phi}\sin{\chi}-4h_y\cos{\omega t}(\sin{\Phi}\cos{\chi}+\varepsilon\cos{\Phi}\sin{\chi}]+\frac{1}{2}(m_z^2+l_z^2+2\varepsilon m_zl_z).\\
\end{aligned}\label{W}
\end{equation}
\end{widetext}

\noindent Here $W_0=\pi[(1+\varepsilon^2)(N_y+N_x)+h_u]M_S^2$; $\Phi,\;\chi$ are the normal coordinates and are given by $\Phi=(\varphi_1+\varphi_2)/2,\;\chi=(\varphi_1-\varphi_2)/2$; $m_z=\xi_1+\xi_2$, $l_z=\xi_1-\xi_2$. Further, the following notations were used: $h_f=(h_1^m+h_2^m)/2$, $h_d=(h_1^m-h_2^m)/2$.
\newline
\newline
\newline
\newline
\newline
\\

The Lagrange density can now be written as \cite{lagrange}
\begin{equation}
\begin{aligned}
\frac{L}{2\pi M_S^2}=&(1+m_z)\left(\frac{d\Phi}{d\tau}+\varepsilon\frac{d\chi}{d\tau}\right)\\
&+l_z\left(\frac{d\chi}{d\tau}+\varepsilon\frac{d\Phi}{d\tau}\right)-W(\Phi,\chi,m_z,l_z),\nonumber
\end{aligned}
\end{equation}
where $\tau=t\cdot\omega_0$, $\,\omega_0=8\pi M_S\mu_B/\hbar$, $\,\Omega=\omega/\omega_0$.
The dynamics of SFi are then described by the following system of equations:
\begin{widetext}
\begin{equation}
\begin{aligned}
&\frac{d^2}{d\tau^2}(\Phi+\varepsilon\chi)=h_y\cos{\Omega\tau}(\cos{\Phi}\cos{\chi}-\varepsilon\sin{\Phi}\sin{\chi})-(h_x+h_f+\varepsilon h_d)\sin{\Phi}\cos{\chi}-[\varepsilon(h_x+h_f)+h_d]\cos{\Phi}\sin{\chi}\\
&\qquad-\frac{1}{2}\Big\{\!\!\left[(1+\varepsilon^2)\left(N_y-N_x\right)+h_u\right]\cos{2\chi}+(1-\varepsilon^2)(\gamma_y-\gamma_x)\!\Big\}\sin{2\Phi}-\varepsilon\left(N_y-N_x+\frac{h_u}{2}\right)\cos{2\Phi}\sin{2\chi},\\
&\frac{d^2}{d\tau^2}(\chi+\varepsilon\Phi)=h_y\cos{\Omega\tau}(\varepsilon\cos{\Phi}\cos{\chi}-\sin{\Phi}\sin{\chi})-(h_x+h_f+\varepsilon h_d)\cos{\Phi}\sin{\chi}-[\varepsilon(h_x+h_f)+h_d]\sin{\Phi}\cos{\chi}\\
&\qquad-\frac{1}{2}\Big\{\!\!\left[(1+\varepsilon^2)\left(N_y-N_x\right)+h_u\right]\cos{2\Phi}-(1-\varepsilon^2)(\gamma_y+\gamma_x)\!\Big\}\sin{2\chi}-\varepsilon\left(N_y-N_x+\frac{h_u}{2}\right)\sin{2\Phi}\cos{2\chi},\\
\label{diff}
\end{aligned}
\end{equation}
\end{widetext}

These equations naturally yield the two ground states in the absence of microwave field ($h_y=0$), with $\Phi=\pi/2$, $\chi=\pm\pi/2$, corresponding to the antiferromagnetic vector values of the bilayer $\textbf{l}=(\pm2,0,0)$.

\begin{figure}[!t]
\includegraphics[width=1\linewidth]{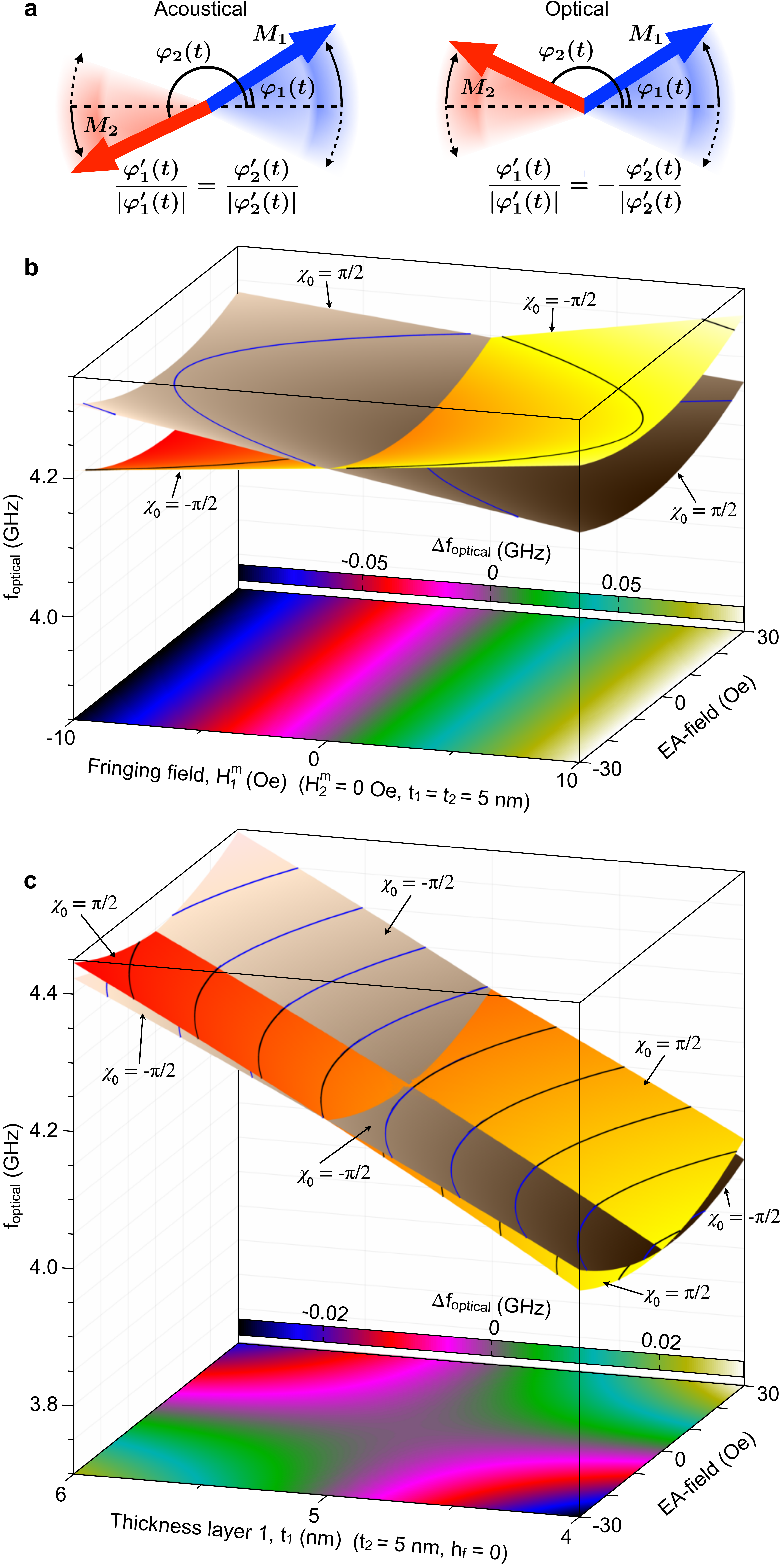}
\caption{(Color online) \textbf{a}, Schematic of the acoustical (in-phase) and optical (out-of-phase) collective resonance modes; the horizontal dotted line represents the easy-axis; the colored shadows depict the individual macrospin oscillations. \textbf{b,c}, The splitting of the optical spin resonance frequency for the two AP states (color scale) as a function of: \textbf{b}, $H_{x}$ and asymmetric biasing (fringing) field, with zero thickness imbalance ; \textbf{c}, $H_{x}$ and thickness imbalance, with zero fringing field. 
}
\label{res_dif}
\end{figure}

Solving Eqs. \ref{diff} for small oscillations about the two ground states yields the frequencies of the collective spin resonances -- the SFi eigen-modes:
\begin{equation}
\begin{aligned}
&\left(\Omega^\pm\right)^{\!2}=N_y\!-\!N_x\!+\!\gamma_x\!+\!h_u\!-\!h_d\sin{\!\chi_0}\!\pm\!\Big\{\!(1\!-\!\varepsilon^2)\gamma_y^2\\
&\qquad+\left[(h_x+h_f)\sin{\chi_0}-\varepsilon(N_y-N_x-\gamma_x)\right]^2\!\Big\}^{\!1/2}.\nonumber
\end{aligned}
\end{equation}
Here "$-$" stands for acoustical (in-phase) and "+" for optical (out-of-phase) resonance frequencies (Fig. \ref{res_dif}a). The criterion for generating spin resonance in both ground states, in terms of $H_x$, is given by
\begin{equation}
\begin{aligned}
h_x=&\frac{g_s(t_1,t_2,H_1^m,H_2^m)}{(4\pi M_S)}\\
&=-h_f-h_d\sqrt{1-\frac{(1-\varepsilon^2)\gamma_y^2}{h_d^2-\varepsilon^2(N_y-N_x-\gamma_x)^2}}.
\label{singlet_eq}
\end{aligned}
\end{equation}
Here all terms are functions of ($t_1,t_2$) or ($H_1^m,H_2^m$). The optical resonance frequencies for the two ground states are plotted in fig. \ref{res_dif}b,c as surfaces in the SFi-asymmetry versus external field parameter space. The bottom plane of figs. \ref{res_dif}b,c are intensity plots of the difference in the optical resonance frequency of the two ground states ($\Delta f_\text{optical}=\Delta\Omega^+\cdot\omega_0/2\pi $), again versus the key asymmetry-field parameters.
\newline

In order to determine the mechanism of resonant SFi switching, we next analyze the behavior of the system for large microwave excitation amplitudes. For this we only consider a fringing field asymmetry (which typically has a stronger effect on the splitting of the optical resonance) and take the thicknesses of the two free layers to be equal ($t_1=t_2\Rightarrow\varepsilon=0$). To simplify the derivations, the external uniform EA-field is set to $h_x=-h_f$, for which the resonance splitting is maximal. 

A hard-axis ($y$-axis) microwave field forces the two macrospins to oscillate out-of-phase, in the optical mode, which results in a modulation of $\chi$. At the same time, $\Phi$ only slightly deviates from its ground state of $\Phi_0=\pi/2$. Therefore, it is a good approximation to take $\varphi=\Phi-\pi/2$ ($|\varphi|\ll1$). As a result, Eqs. (\ref{diff}) can be rewritten as

\begin{equation}
\begin{aligned}
\frac{d^2\chi}{d\tau^2}+2\lambda\Omega_\text{o}\frac{d\chi}{d\tau}-\Omega_\text{o}^2\sin{\chi}\cos{\chi}\qquad&\\
+h_d\cos{\chi}+h\cos{\Omega\tau}\sin{\chi}&=0.
\label{diff_chi3}
\end{aligned}
\end{equation}
\begin{equation}
\begin{aligned}
\frac{d^2\varphi}{d\tau^2}&+\left(\Omega_\text{a}^\textit{eff}\right)^2\cdot\varphi=0,\\
\left(\Omega_\text{a}^\textit{eff}\right)^2&={}\Omega_\text{a}^2-2\left(N_y-N_x+h_u\right)\cdot\overline{\cos^2\chi}\\
&\qquad-h_d\cdot\overline{\sin\chi}+h\cdot\overline{\cos{\Omega\tau}\cos\chi,}
\label{omega_eff}
\end{aligned}
\end{equation}

\noindent provided that $\Omega\approx\Omega_\text{o}$ and $\Omega_\text{a}\ll\Omega_\text{o}$ (which is valid in our case of elliptical particles of aspect ratio approximately 1). $\Omega_\text{a}$ and $\Omega_\text{o}$ are the characteristic frequencies of small amplitude 'acoustical' and 'optical' oscillations in SAF, respectively, and are given by $\Omega_\text{a}=\sqrt{\left(N_y-N_x+h_u\right)-\left(\gamma_y-\gamma_x\right)}$ and $\Omega_\text{o}=\sqrt{\left(N_y-N_x+h_u\right)+\left(\gamma_y+\gamma_x\right)}$.
Under these conditions, the variation in $\varphi$ can be regarded as 'slow' compared to the 'fast' oscillations in $\chi$, in which case averaging the $\chi$- and $\Omega$-terms in equation \ref{omega_eff} over one period is justified (represented by the bar). The $\lambda\Omega_\text{o}$-term in eq. \ref{diff_chi3} is the parameter characterizing the dissipation of energy in the system. Its value is associated with the Landau-Lifshitz-Gilbert damping parameter.

Using Eqs. (\ref{diff_chi3}) and (\ref{omega_eff}), the stability criteria for the ground states can be derived (see supplementary):
\begin{equation}
\begin{aligned}
\frac{h_y}{4\Omega_\text{o}^2}\leq\! A_c\!\sqrt{\!1\!-\!\frac{3A_c^2}{4}}\sqrt{\!\!\left(\!\!\gamma_f\!+\!\frac{h_d\sin{\!\chi_0}}{2\Omega_\text{o}^2}\!\right)^{\!\!2}\!\!+\!\lambda^2\!\left(\!1\!-\!\frac{A_c^2}{4}\!\right)},\nonumber
\end{aligned}
\end{equation}
where $A_c$ is the critical oscillation amplitude derived in the supplementary, $\gamma_f=(\Omega-\Omega_c)/\Omega_c$, with $\Omega_c=\sqrt{\Omega_\text{o}^2(1-3A_c^2/4)/(1+5A_c^2/4)}$. The above result takes into account that $|\gamma_f|\ll1$.  

\begin{figure}
\includegraphics[width=1\linewidth]{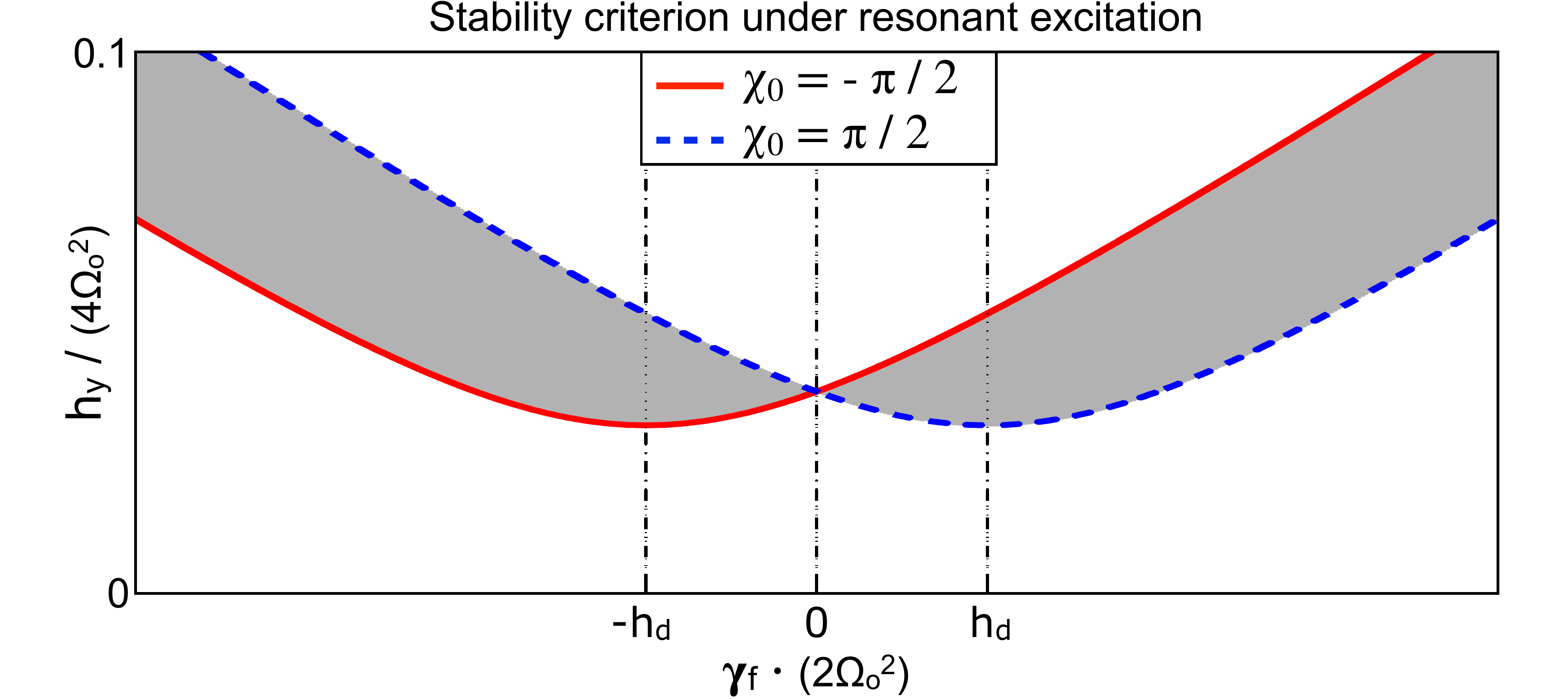}
\caption{(Color online) Stability diagram versus frequency-amplitude for the SFi ground states under resonant excitation. The curves separate the regions of stability and instability (below and above). The greyed out area shows the frequency-amplitude region where only one of the two states is stable, into which the system switches for given excitation. $\Omega_\text{o}$ is the optical center frequency (that of an ideal SAF), $h_y$ the amplitude of the excitation, $\gamma_f=(\Omega-\Omega_c)/\Omega_c\Rightarrow\gamma_f\cdot4\Omega_\text{o}^2\propto\Omega-\Omega_c$, $\Omega$ the excitation frequency.
\label{theory}}
\end{figure}

Thus, in the presence of biasing field asymmetry, the stability regions for the two ground states split, and can fully separate such that microwave frequency-amplitude areas exist where only one states is stable, as illustrated in Fig. \ref{theory}. If resonantly excited within these areas (shown in grey), the system switches into the given stable AP-state and remains in it, now off-resonance with the excitation. The separation of the minima in Fig. \ref{theory} is the splitting in the optical eigen-mode of the system and is given by $\gamma_f=h_d/\Omega_\text{o}^2$.

The separation of the respective optical spin resonance frequencies is due to the difference in the effective biasing field acting on the individual layers. If the amplitude of the excitation is large enough and the frequency is chosen in resonance with a given ground state, switching into this state occurs, while the reverse switching is off-resonance and therefore disallowed energetically.

Interestingly, although the oscillation of the two macrospins is of optical type (out-of-phase), the switching occurs acoustically (in-phase rotation). The dipole-coupling is too strong for the moments to cross passed each other under the relatively weak microwave resonant fields used in this work. The switching trajectory is thus an in-phase rotation with superposed out-of-phase oscillations of smaller amplitude, with the energy effectively transferred from the pumped optical oscillations into a large-angle acoustical rotation. As a result, the expected fastest switching time is the half-period of the acoustical resonance frequency (corresponding to approximately 500 ps for our experimental geometry). For still faster resonant switching one would aim in the SFi-device design at maximizing the acoustical resonance frequency while maintaining a suitable size splitting in the optical resonance frequency; for example, by increasing the aspect ratio of the nanomagnets.

\begin{figure}
\includegraphics[width=1\linewidth]{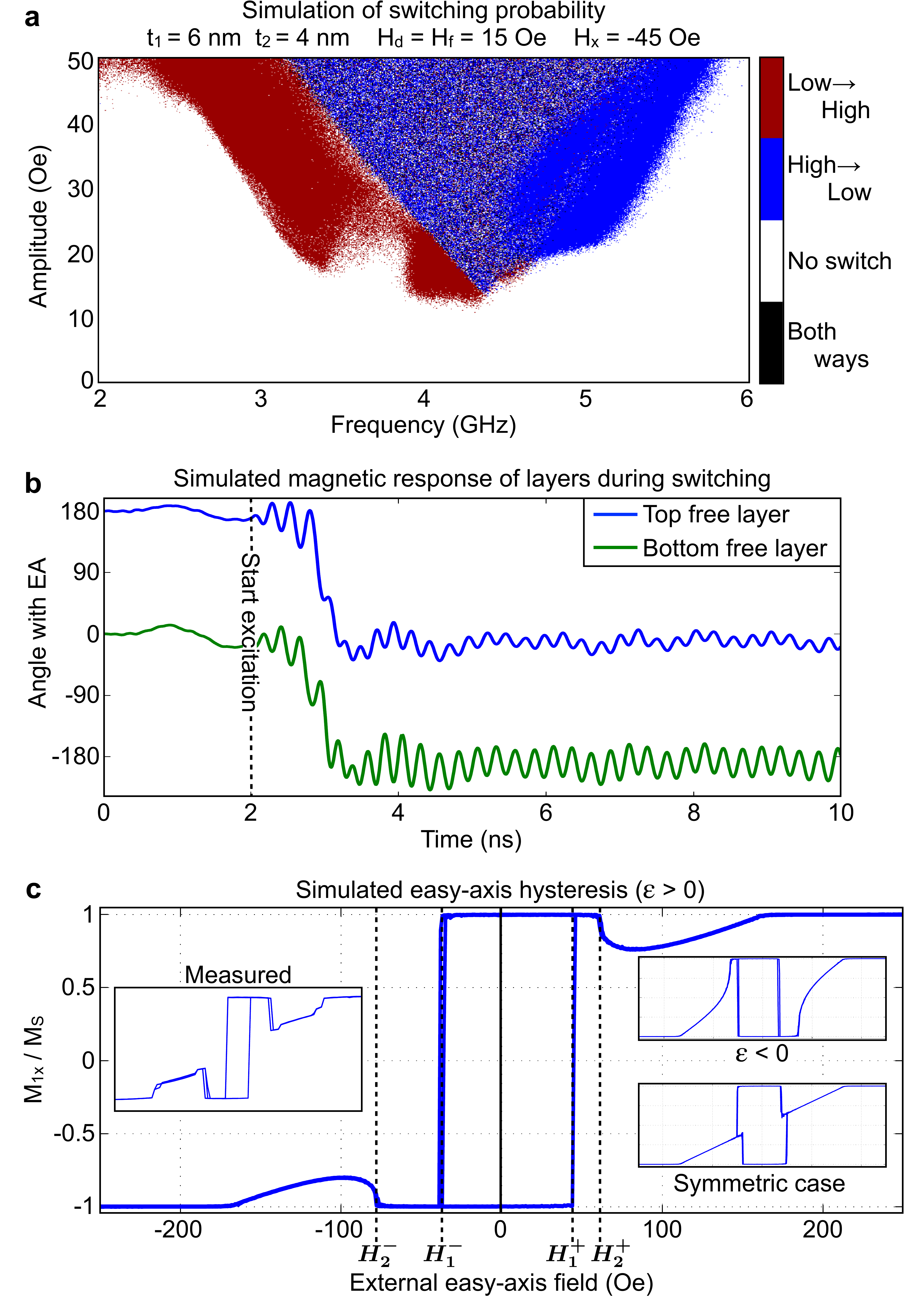}
\caption{(Color online) \textbf{a}, Switching probability map for the sample parameters given in the figure, showing a clear separation of the two AP states due to a biasing field and thickness asymmetry. \textbf{b}, Time trace of the two macrospins in the bilayer during a switching event, showing that the switching occurs via an in-phase rotation (acoustical mode), with superposed optical oscillations of smaller amplitude. \textbf{c}, Simulated easy-axis hysteresis for a sample with a thickness imbalance and biasing field asymmetry for $\varepsilon > 0$ (main panel), corresponding to our typical experimental configuration (left inset), and for $\varepsilon < 0$ (top-right inset). In contrast, an ideal, symmetric spin-flop bilayer shows a strictly anti-symmetric response (bottom-right inset; simulated).
\label{sim}}
\end{figure}

\section{Macrospin simulations}

\noindent The macrospin simulations in this section solve the LLG equations for the system in the time domain, taking into account all the external and internal forces acting on the magnetic moments of the two layers, without the small-signal assumptions used in obtaining the analytical results. The simulation model is an extension of our previous work on the energetics of the SAF system \cite{stat_sw3,SDM_worledge}. The magnetization vectors evolve in time driven by torques of the respective effective magnetic fields, with a suitably small time step of 5 ps. A thickness imbalance and fringing field are the two asymmetry parameters. The simulations include thermal agitation at $T=300$ K.

In the case of no thickness imbalance, with the only asymmetry due to the unequal biasing fields on the individual free layers (from the reference layer built-in into the nanopillar), the numerically simulated switching behavior agrees well with that predicted analytically above. The optical resonance is clearly split, although the magnitude of the splitting is somewhat smaller. 

A more pronounced resonance splitting is obtained by adding a thickness-imbalance to the magnetic asymmetry of the system. Figure \ref{sim}a shows a typical simulated switching map. The area where switching occurs only in one direction is where only one ground state is dynamically stable. The general form of the stability regions simulated here for large-signal microwave excitation agrees well with the analytical results.

Figure \ref{sim}b shows the real-time trace of the magnetization of the individual layers during a switching event. It is clear that the microwave-pumped oscillations are of optical nature but the switching itself is an "acoustical" in-phase rotation. The oscillations of the magnetization in the initial ground state under a resonant excitation of amplitude suitably large to produce switching are, in fact, smaller in the switched-to ground state since this state is now off-resonance with the microwave field. This effect of resonant state selection is a direct consequence of the magnetic asymmetry of SFi.

The full macrospin model further yields important insights into the quasistatic magneto-resistance properties of the junctions. Using Eq. \ref{E} the quasistatic switching ($H_1^\pm$) and spin-flop fields ($H_2^\pm$) can be obtained and used for extracting the asymmetry parameters (individual layers' thicknesses and effective biasing fields) from the measured data. The interrelation between the thicknesses and fringing fields is (in notations of Fig. \ref{sim}(c)):\\
\begin{equation}
\begin{aligned}
\frac{H_i^\pm}{4\pi M_S}&\!=\!\mp(-1)^i|\varepsilon|(N_y\!-\!N_x\!-\!\gamma_x\!)\!-\!h_f\!\\
&\pm\!\sqrt{\!\left(\!N_y\!-\!N_x\!+\!\gamma_x\!+\!h_u\!\pm\!(-1)^{i} \frac{\varepsilon}{|\varepsilon|}h_d\!\right)^{\!2}\!-\!(1\!-\!\varepsilon^2)\gamma_y^2},\\
\\
|\varepsilon|\approx &\frac{h_1^+ \!-\! h_1^- \!-\! h_2^+ \!+\! h_2^-}{4(N_y-N_x-\gamma_x)},\quad h_f\approx\frac{-(h_1^+ \!+\! h_1^- \!+\! h_2^+ \!+\! h_2^-)}{4}.\nonumber
\end{aligned}
\end{equation}

\noindent Here $H_1^\pm=4\pi M_S h_1^\pm $ is the field for which only one AP ground state becomes unstable and $H_2^\pm=4\pi M_S h_2^\pm$ is the spin-flop field, at which both AP ground states become unstable. Note that the actual values of $H_{1,2}^\pm$ are slightly smaller than the analytical values due to thermal fluctuations.

\section{Experiments}

\begin{figure*}[!ht]
\includegraphics[width=1\linewidth]{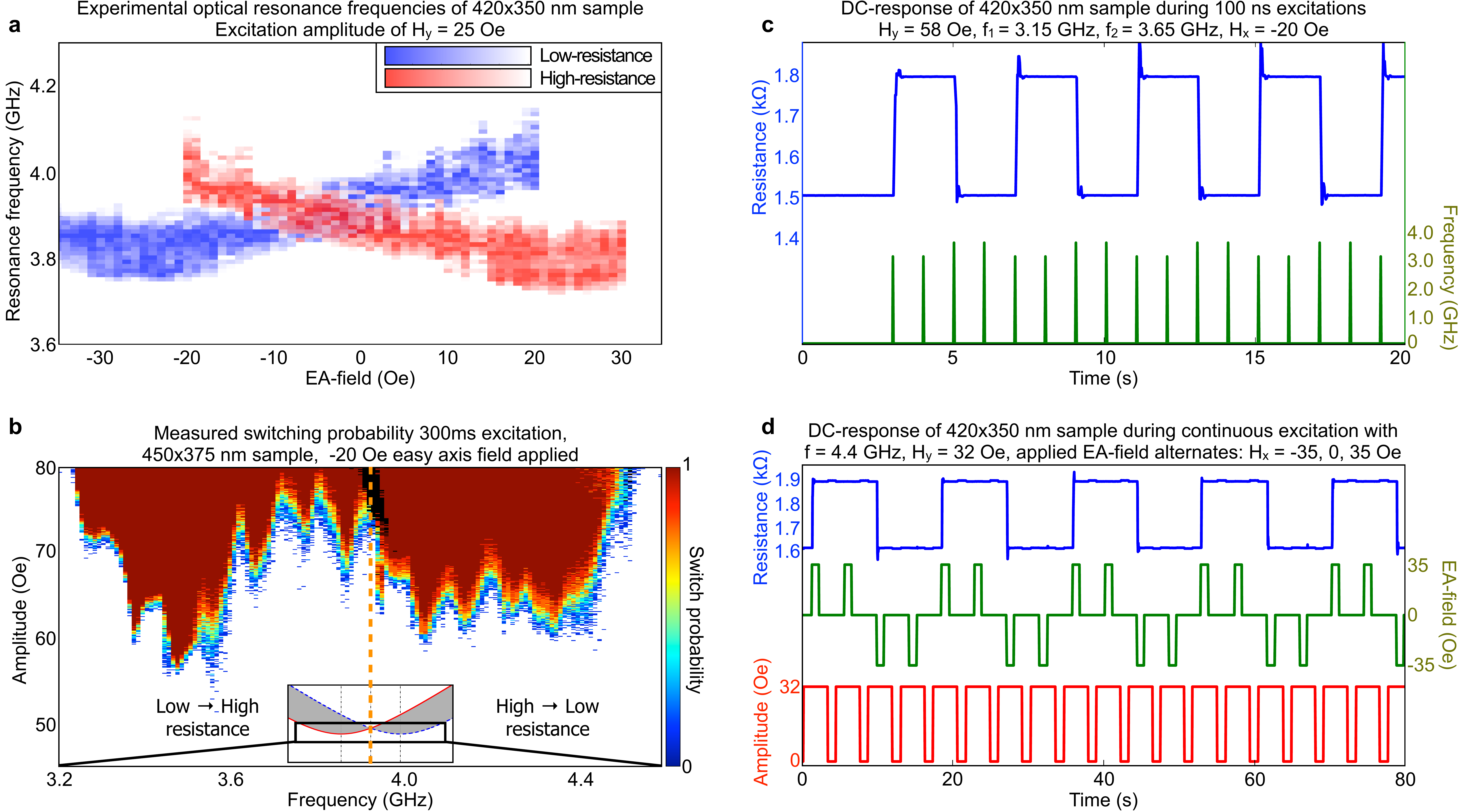}
\caption{(Color online) \textbf{a}, Measured resonance frequencies of a 420x350 nm spin-flop bilayer. The intensity map (red and blue for the two ground states) is normalized to the maximum oscillation amplitude for the given EA-field. \textbf{b}, Switching probability for the two ground states subject to 300 ms long excitation pulses with -20 Oe static EA-field applied. Black color denotes the frequency-amplitude points, for which switching in both directions occurred. The inset shows the section of the analytical switching map (Fig. \ref{theory}) covered in the measurement. \textbf{c}, Real time resistance traces of a sample excited by short microwave pulses of alternating frequencies in resonance with the split optical mode, showing controlled resonant state selection. \textbf{d}, Resonant state selection using resonant microwave pumping of fixed frequency and alternating external EA-field, tuning the two ground states in and out of the optical spin resonance.
\label{fmr}}
\end{figure*}

\subsection{Samples}
The samples used in the experiments were spin-flop type magnetic random access memory (TMRAM) cells with lateral dimensions of 450x375 and 420x350 nm, and elliptical in-plane shape. The stack consists of a magnetically soft SFi separated from a magnetically hard SAF reference layer by a thin Al-O tunnel barrier. The spin-flop bilayer is composed of two dipole coupled permalloy ferromagnets (free layers) with thicknesses $t_1\approx t_2\approx5$ nm (excluding the magnetically 'dead' layers at the film surfaces \cite{dead_layer}) separated by a $\sim$1 nm thin TaN spacer, for which there is no interlayer exchange coupling, $J=0$. Here $t_1$ and $t_2$ are the effective thicknesses of the bottom and top free layer, respectively. The SFM reference layer is nearly ideally flux-closed and is therefore essentially insensitive to external fields of strengths used in the experiments discussed in this paper. The samples used in this work had a weak fringing field from the reference layer acting on the two free layers. The fringing field originated predominantly from the top fixed layer. The resistance of the stack was $\sim$1 k$\Omega$, and the magnetoresistance $\sim$20\%.

\subsection{Measurement setup}
In-plane quasi-static fields were applied using an external toroidal magnet \cite{toroidal_magnet}. High-frequency fields were generated by contacting the on-chip integrated 50 $\Omega$ bit- and word-lines using surface probes with 0-40 GHz bandwidth. The bit- and word-lines were oriented at $\pm$ 45$^{\circ}$ with respect to the the easy-axis of the stack in the so-called "toggling" configuration. A current through the bit- or word-line results in an in-plane magnetic field perpendicular to the line. In the results presented below, the high frequency excitation was applied using the word line only, which was electrically decoupled from the stack. The applied high-frequency fields were AC pulses of desired frequency ranging from 1 MHz to 10 GHz. The AC fields were generated using an Agilent E8247C PSG CW Signal Generator with a bandwidth of 250 kHz - 20 GHz. The wave-duration was controlled using an RF-switch driven by an Agilent 33250A AWG. The shortest possible pulse-length with this setup was 8 ns. The state of the sample was determined by measuring the resistance of the sample and thereby the relative orientation of the magnetization of the bottom free and top fixed layers. To separate the DC-signal from the RF-signal, a bias-tee was used.

\subsection{Experimental results}
\noindent The samples were SFi bilayers of approximately 5 nm thick Permalloy, separated by 1 nm thick TaN, integrated into approximately 350x400 nm in-plane elliptical nanopillars, also containing an Al-O read-out junction with a nominally flux-closed magnetic reference layer. The microwave excitation was applied using a closely spaced field-line making a 45$^\circ$ angle with the SFi easy axis. The ferromagnetic resonance frequencies were measured using transport spectroscopy, where the junction resistance versus frequency showed a maximum or a minimum for the low-resistance or high-resistance ground state, respectively. The resistance was measured while slowly sweeping the frequency of a continuous-wave excitation of fixed amplitude (typically 25 Oe). Figure \ref{fmr}a shows the measured resonance frequencies for the two AP ground states of a SFi with $M_0<0$, $H_1^m \approx 7$ Oe, $H_2^m \approx 3$ Oe and $t_1-t_2\approx0.9$ nm (values deduced from quasistatic EA-hysteresis as shown in fig \ref{sim}c). The splitting of the optical spin resonance at $\approx 4$ GHz is in excellent agreement with the predicted behavior (Fig. \ref{theory}).  

Figure \ref{fmr}a shows the measured oscillation amplitude as a function of the excitation frequency and applied bias field. The amplitude is normalized with respect to the maximum value for each bias field. This diagram clearly shows that both field-controlled and frequency-controlled switching is possible.

Frequency-amplitude-switching maps were measured for both AP states (low- to high-resistance and high- to low-resistance) for different excitation pulse durations and static EA-fields. Figure \ref{fmr}b shows the switching maps for both AP states of the soft bilayer under 300 ms pulses of resonant excitation, with $H_{x}=-20$ Oe. Switching at frequencies lower than 3.9 GHz only occurs from the low- to high-resistance AP state ($f_{L\rightarrow H}$), while switching at above 3.9 GHz only occurs from the high- to low-resistance state ($f_{H\rightarrow L}$). Shortening the duration of the excitation pulse requires a higher amplitude for switching the junction, at the same time leading to an increase in the error-rate likely due to a more non-linear and more non-uniform process for higher-amplitude pumping. Nevertheless, even for short excitations, the optical resonance is clearly split and amplitude-frequency regions exist where the switching is strictly one-directional. The weaker higher-order oscillations in the probability maps of Fig. \ref{fmr}b are due to secondary spin-wave excitations in our nanoparticles that are somewhat bigger than the true single-domain limit, which we also observe in our micromagnetic simulations as a small superposition on the well-defined macrospin behavior.

Figure \ref{fmr}c shows a realtime trace of the junction resistance during the application of a sequence of microwave pulses of 100 ns duration, of frequency alternating between the split optical spin-resonance peaks. A pair of pulses is sequentially applied at each frequency, demonstrating the strictly unidirectional character of resonant switching at a given frequency (for a given AP state of the SFi). 

Figure \ref{fmr}d shows a realtime resistance trace (blue) under continuous microwave excitation of fixed frequency near the centre of the optical resonance, with the EA-field of amplitude sufficient for tuning the two AP ground states in and out of the resonance. Between the EA-pulses the excitation is turned off shortly to confirm the state the sample is in (microwave amplitude shown in red; EA-field in green). The EA-field pulses were generated with an external electromagnet, so their shortest duration was 300 ms. The data show that the field-controlled resonance state selection is highly reliable.

Both measurements were repeated many times with the resulting error-rate smaller than $10^{-3}$ (1000 measurements were performed without any error). The effect is thus essentially deterministic for the sample and frequency-field parameters used, and still is expected to improve (e.g., in terms of speed) on scaling down the sample size.

\section{Conclusions}
\noindent  We show how external microwave and static biasing fields can be used to tune the energetics and dynamics of a spin-flop system between symmetric and asymmetric behavior. We demonstrate controlled state selection in a synthetic ferrimagnet using a frequency modulated resonant microwave field or a uniform external biasing field. This effect is explained analytically using a small-signal analysis of the SFi spin dynamics, predicting a tunable splitting of the optical spin resonance, which is in good agreement with our numerical arbitrary signal-strength simulations. The effect is fast, robust, and offers a new way of magnetic switching of spin-flop nanodevices used in such large scale applications as magnetic random access memory.

Our analysis shows that the switching process is fast, $\sim 1/2f_a \sim 500$ ps, and proceeds through an in-phase (acoustical) spin rotation with superposed out-of-phase (optical) spin oscillations. Experimentally we demonstrate essentially error-free controlled switching down to 100 ns in pulse duration of the resonant microwave excitation. Smaller, more single-domain-like samples (sub-100 nm range), with optimized magnetic asymmetry and increased aspect ratio are expected to show significantly faster resonant switching due to reduced generation of unwanted spin-wave modes and higher acoustical frequencies.



\begin{widetext}
\newpage
\section*{\large{Supplementary: Controlled resonant switching of a synthetic ferrimagnet}}
\end{widetext}

\renewcommand\theequation{S\arabic{equation}}
\setcounter{equation}{0}
\section*{Macrospin theory}

\noindent Considering the free layers of the synthetic ferrimagnet (SFi) as single-domain particles, with the the z-axis out-of-plane, the EA and a uniaxial anisotropy field along the x-axis, the total magnetic energy ($E$) of the two free layers can be written as \cite{stat_sw3}:
\begin{equation}
\begin{aligned}
\frac{E}{2\pi M_S^2}=&\sum_{i,j=1;i\neq j}^2\!\!\!{V_i\left\{N_{ix}m_{ix}^2+(N_{iy}+h_{u})m_{iy}^2+N_{iz}m_{iz}^2\right.}\nonumber\\
&+\gamma_{jx}m_{1x}m_{2x}+\gamma_{jy} m_{1y}m_{2y}+\gamma_{jz} m_{1z}m_{2z}\label{E}\nonumber\\
&{\left.-2(h_x+h_{ix}^m)m_{ix}-2h_y\cos{(\omega t)}m_{iy}\right\}},\nonumber
\end{aligned}
\end{equation}

\noindent where $M_S$ is the saturation magnetisation (assumed same in both layers); $V_i=\pi abt_i/4$ the volume of layer $i$ with $a,\,b$ and $t_i$ the length, width and thickness of that layer, respectively; $N_{i\alpha}$ the demagnetising factors of layer $i$ in the $\alpha$-dimension with $\sum_\alpha N_{i\alpha}=1$ and $N_{i\{x,y\}}=n_{\{x,y\}}t_i/b$ where $n_{\{x,y\}}$ are the reduced factors \cite{worledge2}; $\textbf{m}_i$ the cartesian unit magnetisation vector of the $i$-th layer; $h_u=H_u/4\pi M_S$ the normalised intrinsic uniaxial anisotropy field along the easy-axis; $h_y=H_{y}/4\pi M_S$ the normalised amplitude of the applied AC-field in the y-direction; $\omega$ the frequency of the AC-field; $h_x=H_{x}/4\pi M_S$ the normalised external easy-axis DC-field; $h_i^m=H_i^m/4\pi M_S$ the normalised fringing field acting on layer $i$ in the $x$-direction; $\gamma_{i\alpha}=r_\alpha N_{i\alpha}$ the inter-layer exchange field from layer $i$ acting on the other layer in the $\alpha$-direction \cite{dipole_couple}. In our case $\gamma_{i\alpha}$ is solely of dipolar nature. 
Since in a thin magnetic film the in-plane demagnetising factors are proportional to their thickness, they can be rewritten as

\begin{equation}
N_{i\{x,y\}}\!=\!\left[1-(-1)^i\,\varepsilon\right]N_{\{x,y\}},\quad\!\! N_{iz}\!=\!1-N_{ix}-N_{iy},\nonumber
\end{equation}
where
\begin{equation}
N_{\{x,y\}}=\frac{\left(t_{1}+t_{2}\right)n_{\{x,y\}}}{2b},\quad\!\! \varepsilon=\frac{t_1-t_2}{t_1+t_2}.\nonumber
\end{equation}
The same can be done for the volume and interlayer exchange field of the individual layers:
\begin{equation}
V_i=\left[1-(-1)^i\varepsilon\right]V/2;\quad\!\!\!\gamma_{i\{x,y\}}=\left[1-(-1)^i\varepsilon\right]\gamma_{\{x,y\}},\nonumber
\end{equation}

For the description of the dynamics of the magnetisation of the system, it is useful to convert the magnetisation into polar coordinates:
\begin{equation}
\textbf{m}_i=(\,\cos{\varphi_i}\,\sin{\theta_i},\quad\sin{\varphi_i}\,\sin{\theta_i},\quad\cos{\theta_i}\,),\nonumber
\end{equation}
Since in the used geometry the thicknesses are much smaller than the lateral dimensions and the sample is longer along the x-dimension than the y-dimension, the following holds:
\begin{equation}
N_{ix}<N_{iy}\ll N_{iz}\approx 1;\quad\theta_i=\pi/2+\xi_i,\:\: |\xi_i|\ll1,\nonumber
\end{equation}
Hence, $\cos(\theta_i)\approx-\xi_i$. The terms $r_z$, $\xi_1$, $\xi_2$ are minimal and therefore $\gamma_z\xi_i\xi_j$ can be neglected:
\begin{widetext}
\begin{equation}
\begin{aligned}
\frac{W-W_0}{\pi M_S^2}=&-(1+\varepsilon^2)\left(N_y-N_x+\frac{h_u}{1+\varepsilon^2}\right)\cos{2\Phi}\cos{2\chi}+2\varepsilon\left(N_y-N_x+\frac{h_u}{2}\right)\sin{2\Phi}\sin{2\chi}\\
&\quad-(1-\varepsilon^2)\left[(\gamma_y-\gamma_x)\cos{2\Phi}+(\gamma_y+\gamma_x)\cos{2\Phi}\right]-4(h_x+h_f+\varepsilon h_d)\cos{\Phi}\cos{\chi}\\
&\quad+4[\varepsilon(h_x+h_f)+h_d]\sin{\Phi}\sin{\chi}-4h_y\cos{\omega t}(\sin{\Phi}\cos{\chi}+\varepsilon\cos{\Phi}\sin{\chi}]+\frac{1}{2}(m_z^2+l_z^2+2\varepsilon m_zl_z),\\
\end{aligned}\label{W}
\end{equation}
\end{widetext}
\noindent where $W_0=\pi[(1+\varepsilon^2)(N_y+N_x)+h_u]M_S^2$; $\Phi,\;\chi$ are the normal coordinates and are given by $\Phi=(\varphi_1+\varphi_2)/2$, $\chi=(\varphi_1-\varphi_2)/2$; $m_z=\xi_1+\xi_2$, $l_z=\xi_1-\xi_2$, here $m_z$ and $l_z$ are the normalised z-components of the magnetisation vectors of the resulting pair of layers. Further the following relations were used: $h_f=(h_1^m+h_2^m)/2$, $h_d=(h_1^m-h_2^m)/2$.

The Lagrange density is then given by \cite{lagrange}:
\begin{equation}
L=\frac{1}{2}\sum_{i=1}^{2}\frac{\hbar M_S}{2\mu_B}(1-\cos{\theta_i})\frac{\partial\varphi_i}{\partial t}(1-(-1)^i\varepsilon)-W.\nonumber
\end{equation}
Hence,
\begin{equation}
\begin{aligned}
\frac{L}{2\pi M_S^2}=\,&(1+m_z)\left(\frac{d\Phi}{d\tau}+\varepsilon\frac{d\chi}{d\tau}\right)\\
&+l_z\left(\frac{d\chi}{d\tau}+\varepsilon\frac{d\Phi}{d\tau}\right)-W(\Phi,\chi,m_z,l_z),\nonumber
\end{aligned}
\end{equation}
where $\tau=t\cdot\omega_0$, $\,\omega_0=8\pi M_S\mu_B/\hbar$, $\,\Omega=\omega/\omega_0$.
Therefore, the dynamics of the system are described by:
\begin{widetext}
\begin{equation}
\begin{aligned}
&\!\!\!\frac{d^2}{d\tau^2}(\Phi+\varepsilon\chi)=h_y\cos{\Omega\tau}(\cos{\Phi}\cos{\chi}-\varepsilon\sin{\Phi}\sin{\chi})-(h_x+h_f+\varepsilon h_d)\sin{\Phi}\cos{\chi}-[\varepsilon(h_x+h_f)+h_d]\cos{\Phi}\sin{\chi}\\
&\qquad-\frac{1}{2}\Big\{\!\!\left[(1+\varepsilon^2)\left(N_y-N_x\right)+h_u\right]\cos{2\chi}+(1-\varepsilon^2)(\gamma_y-\gamma_x)\!\Big\}\sin{2\Phi}-\varepsilon\left(N_y-N_x+\frac{h_u}{2}\right)\cos{2\Phi}\sin{2\chi},\\
&\!\!\!\frac{d^2}{d\tau^2}(\chi+\varepsilon\Phi)=h_y\cos{\Omega\tau}(\varepsilon\cos{\Phi}\cos{\chi}-\sin{\Phi}\sin{\chi})-(h_x+h_f+\varepsilon h_d)\cos{\Phi}\sin{\chi}-[\varepsilon(h_x+h_f)+h_d]\sin{\Phi}\cos{\chi}\\
&\qquad-\frac{1}{2}\Big\{\!\!\left[(1+\varepsilon^2)\left(N_y-N_x\right)+h_u\right]\cos{2\Phi}-(1-\varepsilon^2)(\gamma_y+\gamma_x)\!\Big\}\sin{2\chi}-\varepsilon\left(N_y-N_x+\frac{h_u}{2}\right)\sin{2\Phi}\cos{2\chi},\\
\label{diff}
\end{aligned}
\end{equation}
\begin{equation}
m_z=2\frac{d\Phi}{d\tau};\qquad l_z=2\frac{d\chi}{d\tau}.\nonumber\\
\end{equation}

\end{widetext}
From these equations it is also easy to determine that the ground states in the absence of the oscillating y-field ($h_y=0$) is given by $\Phi=\pi/2$, $\chi=\pm\pi/2$ which corresponds to the antiferromagnetic vector values $\textbf{l}=(\pm2,0,0)$.

\subsection{Resonance frequency}
\noindent To switch the SFi resonantly between its two ground states, it is necessary to determine the resonance-frequencies of the ground states. If the amplitude of the oscillating field is small ($|h_y|\ll1$), then the magnetic moments of the SFi deviate only minimal from their ground state:
\begin{equation}
\Phi=\frac{\pi}{2}+\Phi_1;\quad \chi=\chi_0+\chi_1;\quad\chi_0=\pm\frac{\pi}{2};\quad|\Phi_1|,|\chi_1|\ll1
\label{small_angle}
\end{equation}

Substituting the relations of (\ref{small_angle}) into Eqs. (\ref{diff}) gives:
\begin{align}
&\frac{d^2\Phi_1}{d\tau^2}+\Omega_{11}^2\Phi_1+\varepsilon\frac{d^2\chi_1}{d\tau^2}+\Omega_{12}^2\chi_1=\varepsilon h_y\cos{\Omega\tau}\sin{\chi_0},\nonumber\\
&\frac{d^2\chi_1}{d\tau^2}+\Omega_{22}^2\chi_1+\varepsilon\frac{d^2\Phi_1}{d\tau^2}+\Omega_{21}^2\Phi_1= h_y\cos{\Omega\tau}\sin{\chi_0},\nonumber\\
\label{diff_small}\\
&\Omega_{ii}^2=(1+\varepsilon^2)(N_y-N_x)+(1-\varepsilon^2)(\gamma_x+(-1)^i\gamma_y)\nonumber\\
&\qquad\qquad+\left[\varepsilon(h_x+h_f)+h_d\right]\sin{\chi_0},\nonumber\\
&\Omega_{ij}^2=2\varepsilon(N_y-N_x+h_u/2)-(h_x+h_f+\varepsilon h_d)\sin{\chi_0},\nonumber¤
\end{align}
\noindent where $i\neq j$. The eigenfrequencies or resonance-frequencies of a system of equations as in Eqs. (\ref{diff_small}) is easy to solve, and is given by
\begin{equation}
\begin{aligned}
&\left(\Omega^\pm\right)^{\!2}=N_y\!-\!N_x\!+\!\gamma_x\!+\!h_u\!-\!h_d\sin{\!\chi_0}\!\pm\!\Big\{\!(1\!-\!\varepsilon^2)\gamma_y^2\\
&\qquad+\left[(h_x+h_f)\sin{\chi_0}-\varepsilon(N_y-N_x-\gamma_x)\right]^2\!\Big\}^{\!1/2}\nonumber
\end{aligned}
\end{equation}
where "$-$" stands for acoustical (in-phase) and "+" for optical (out-of-phase) resonance frequencies. Hence, the criterion for $H_x$ for degenerate resonance in both states is given by
\begin{equation}
\begin{aligned}
h_x=-h_f-h_d\sqrt{1-\frac{(1-\varepsilon^2)\gamma_y^2}{h_d^2-\varepsilon^2(N_y-N_x-\gamma_x)^2}},\nonumber
\label{singlet_eq}
\end{aligned}
\end{equation}
where all the terms are functions of ($t_1,t_2$) or ($H_1^m,H_2^m$).
The difference in optical frequency of the two ground states ($\Delta f_\text{optical}=\Delta\Omega^+\cdot\omega_0/2\pi $) can be approximated by the linearisation of the frequencies around $\varepsilon=0$ and $h_d=0$:
\begin{equation}
\Delta\Omega^\pm\approx\frac{h_d\pm\varepsilon(h_x\!+\!h_f)\frac{(N_y-N_x-\gamma_x)}{\sqrt{\gamma_y^2+(h_x+h_f)^2}}}{\sqrt{N_y\!-\!N_x\!+\!\gamma_x\!+\!h_u\pm\!\sqrt{\gamma_y^2+(h_x\!+\!h_f)^2}}}\nonumber
\end{equation}

\subsection{Resonant stability}
To determine the switching mechanism for the SFi using resonant excitation we analyse the behaviour and stability of the system for large excitation amplitudes. For this we only consider a fringing field asymmetry (which has the largest contribution to the splitting of resonance) and take the thicknesses of the two free layers to be equal ($t_1=t_2\Rightarrow\varepsilon=0$). To simplify the derivation, the external DC field is set to $h_x=-h_f$ (This is the external field at which the splitting is maximal).

When an external AC-field of suitably high frequency is applied along the hard-axis ($y$-axis) of the elliptical magnetic particles, the magnetic moments begin to oscillate in the optical mode, which results in a modulation of $\chi$. At the same time, $\Phi$ only deviates slightly from its ground state $\Phi_0=\pi/2$. Therefore, it is a good approximation to take $\varphi=\Phi-\pi/2$ ($|\varphi|\ll1$). Taking all the above assumptions in consideration, Eqs. (\ref{diff}) can be rewritten as

\begin{equation}
\begin{aligned}
&\frac{d^2\varphi}{d\tau^2}+\big[\Omega_\text{a}^2-2(N_y-N_x+h_u)\cos^2{\!\chi}\\
&\qquad\qquad-h_d\sin{\!\chi}+h_y\cos{\Omega\tau}\cos{\!\chi}\big]\varphi=0,
\label{diff_phi2}
\end{aligned}
\end{equation}
\begin{equation}
\frac{d^2\chi}{d\tau^2}-\Omega_\text{o}^2\sin{\!\chi}\cos{\!\chi}+h_d\cos{\!\chi}+h_y\cos{\Omega\tau}\sin{\!\chi}=0,
\label{diff_chi2}
\end{equation}

\noindent with $\Omega_\text{a}$ and $\Omega_\text{o}$ the characteristic frequency of small amplitude 'acoustical' and 'optical' oscillations, respectively, in a symmetric system and are given by $\Omega_\text{a}=\sqrt{\left(N_y-N_x+h_u\right)-\left(\gamma_y-\gamma_x\right)}$ and $\Omega_\text{o}=\sqrt{\left(N_y-N_x+h_u\right)+\left(\gamma_y+\gamma_x\right)}$.
Provided that $\Omega_\text{a}\ll\Omega_\text{o}$, (which is valid for elliptical particles with aspect ratio of approximately 1) the variation in $\varphi$ can be regarded as 'slow' compared to the 'fast' oscillations of $\chi$. If the frequency of the applied external field is near the optical resonance frequency $\Omega_\text{o}$, then averaging the $\chi$- and $\Omega$-terms in Eq. (\ref{diff_phi2}) over one period is justified and yields
\begin{equation}
\frac{d^2\varphi}{d\tau^2}+\left(\Omega_\text{a}^\textit{eff}\right)^2\cdot\varphi=0,
\label{diff_phi3}
\end{equation}
\begin{equation}
\begin{aligned}
\left(\Omega_\text{a}^\textit{eff}\right)^2={}&\Omega_\text{a}^2-2\left(N_y-N_x+h_u\right)\cdot\overline{\cos^2\chi}\\
&\quad-h_d\cdot\overline{\sin\chi}+h_y\cdot\overline{\cos{\Omega\tau}\cos\chi,}
\label{omega_eff}
\end{aligned}
\end{equation}

where the bar stands for averaging over one period of the 'fast' oscillation. Using Eq. (\ref{diff_phi3}) the stability of different magnetic configurations resulting from the action of the AC-field excitation can be determined. Whenever $\Omega_\text{a}^\textit{eff}$ becomes complex, the state of the particle becomes unstable, which can result in switching.

To evaluate Eq. (\ref{diff_chi2}) for external excitations at the optical frequency we add a phenomenological term describing the dissipative processes in the system:
\begin{align}
\frac{d^2\chi}{d\tau^2}+2\lambda\Omega_\text{o}\frac{d\chi}{d\tau}-\Omega_\text{o}^2\sin{\chi}\cos{\chi}&\nonumber\\
+h_d\cos{\chi}+h_y\cos{\Omega\tau}\sin{\chi}&=0.
\label{diff_chi3}
\end{align}
Here $\lambda\Omega_\text{o}$ is the parameter characterising the dissipation of energy in the system. Its value can be associated with a dissipative Gilbert parameter.

Starting in the ground states ($\Phi=\pi/2$ and $\chi=\pm\pi/2$) and turning on the AC-field results in periodic oscillations of $\chi$, which will be described using a new variable $f$, connected with $\chi$ by the following relation:
\begin{equation}
\chi=\chi_0+2\arctan(f)\label{chi_f},\quad\chi_0=\pm\pi/2
\end{equation}
Here and further on the upper sign corresponds to the state in which the magnetic moments fluctuate in the vicinity of $\vec l=\left(2,0,0\right)$, and the lower sign to fluctuations in the vicinity of $\vec l=\left(-2,0,0\right)$.

\noindent Substituting Eq. (\ref{chi_f}) into Eq. (\ref{diff_chi3}) results in a fourth order differential equation:

\begin{equation}
\begin{aligned}
&\left(1+f^2\right)\left(\frac{d^2}{d\tau^2}+2\lambda\Omega_\text{o}\frac{d}{d\tau}+\Omega_\text{o}^2-h_d\sin{\chi_0}\right)f\\
&\!-\!2\!\left[\!\left(\!\frac{df}{d\tau}\!\right)^{\!\!2}\!\!+\Omega_\text{o}^2f^2\right]\!\!f\!+\!\frac{h_y}{2}\sin{\chi_0}\cos{\Omega\tau}\!\left(1\!-\!f^4\right)\!=0.
\label{diff_f}
\end{aligned}
\end{equation}
To reduce the degree of nonlinear terms in Eq. (\ref{diff_f}), the accuracy of the theory is limited to
\begin{equation}
f^4\ll1.\nonumber
\end{equation}
This requirement is not too restrictive and allows to consider nonlinear oscillations with high amplitude. Following the standard procedure for finding the resonance in a nonlinear system, such as Eq. (\ref{diff_f}), the solution is represented in the form
\begin{equation}
f=A\cos{\left(\Omega\tau+\delta\right)}+\zeta\label{f},
\end{equation}
with $A$ the amplitude of the main harmonic, $\delta$ the phase-shift with respect to the applied field, and $\zeta$ an additional term taking into account anharmonic corrections.

When the external excitation frequency $\Omega$ is close to the resonance frequency, then the main harmonic oscillations dominate and $|\zeta|\ll A$, in which case the anharmonic corrections can be neglected.
Substituting Eq. (\ref{f}) into Eq. (\ref{diff_f}) and only considering the terms proportional to the main harmonics, ($\sin{(\Omega\tau)}$ and $\cos{(\Omega\tau)}$), the resulting amplitude and phase-shift of the oscillations become:

\begin{equation}
\begin{aligned}
\frac{h_y^2}{4A^2}\approx&\left[\Omega^2\!\left(\!1\!+\!\frac{5A^2}{4}\!\right)\!-\left(\!\Omega_\text{o}^2\!-\!h_d\sin{\!\chi_0}\!\right)\!\left(\!1\!-\!\frac{3A^2}{4}\!\right)\!\right]^{\!2}\\
&\qquad+\left(2\lambda\Omega\Omega_\text{o}\right)^2\!\left(\!1+\frac{A^2}{4}\right)^{\!\!2},
\label{amp_f}
\end{aligned}
\end{equation}
\begin{equation}\tan{\delta}=\frac{2\lambda\Omega\Omega_\text{o}(1+A^2/4)}{\Omega^2(1\!+\!5A^2\!/4)\!-\!(\Omega_\text{o}^2\!-\!h_d\sin{\!\chi_0})(1\!-\!3A^2\!/4)}.
\label{delta_f}
\end{equation}
From these equations it follows that at the excitation frequency of $\Omega\!=\!\sqrt{(\Omega_\text{o}^2\!-\!h_d\sin{\chi_0})(1\!-\!3A^2/4)/(1\!+\!5A^2/4)}$ the system is in resonance when the dissipation is weak. In this case the amplitude reaches a maximum $A_{max}$ defined by $A_{max}(1+A_{max}^2/4)=h_y/4\lambda\Omega\Omega_\text{o}$.

Substituting Eq. (\ref{f}) into (\ref{chi_f}) into (\ref{omega_eff}) and taking the appropriate averages gives the effective 'acoustic' resonance frequency and with that the stability of the ground states:
\begin{widetext}
\begin{equation}
\left(\,\Omega_\text{a}^\textit{eff}\,\right)^2=\Omega_\text{a}^2-\frac{(N_y-N_x+h_u)4A^2}{(1+A^2)^{3/2}}-h_d\sin{\chi_0}\left(\frac{2}{\sqrt{1+A^2}}-1\right)-2\cos{\delta}\sin{\chi_0}\frac{h_y}{A}\left(1-\frac{1}{\sqrt{1+A^2}}\right).
\label{omega_eff2}
\end{equation}
\end{widetext}
When the oscillation amplitude $A$ substantially increases, the right hand side of Eq. (\ref{omega_eff2}) becomes negative, which makes the respective state unstable. The difference in the resonance frequency for the two ground states $\vec l=(2,0,0)$ and $\vec l=(-2,0,0)$ is proportional to the asymmetry in the fringing field ($h_d$). Therefore, it is possible to determine the region in the frequency-amplitude phase space, in which one state is unstable while the other is stable. It is this asymmetry-induced selectivity in, for example, frequency for a given amplitude that allows to uniquely select a given AP state of the SFi.

To determine the frequency-amplitude range where only one of the states is stable, the critical amplitude at which $\Omega_\text{a}^\textit{eff}$ becomes imaginary needs to be determined for the the two AP states. To do this, Eq. (\ref{omega_eff2}) needs to be solved for the case where $\Omega_\text{a}^\textit{eff}=0$. Assuming that the fringing field difference and the AC-field amplitude are small, then the last two terms in (\ref{omega_eff2}) can be neglected, and the critical amplitude $A_c$ becomes:
\begin{equation}
4A_c^2\left(1+A_c^2\right)^{-3/2}=\frac{\left(N_y-N_x+h_u\right)-\left(\gamma_y-\gamma_x\right)}{\left(N_y-N_x+h_u\right)},\nonumber
\end{equation}
which reduces to a cubic equation and is straightforward to solve. For our samples the equation has three real solutions, of which one, $A_c^2=0.118$, satisfies the criterion of applicability, $A_c^4\ll1$.

Once $A_c$ is determined, Eqs. (\ref{amp_f}) and (\ref{delta_f}) can be used to determine the frequency-amplitude values for which each state changes between stable and unstable. Hence, a ground state is only stable if
\begin{align}\frac{h_y}{4\Omega_\text{o}^2}\leq\! A_c\!\sqrt{\!1\!-\!\frac{3A_c^2}{4}}\sqrt{\!\!\left(\!\!\gamma_f\!+\!\frac{h_d\sin{\!\chi_0}}{2\Omega_\text{o}^2}\!\right)^{\!\!2}\!\!+\!\lambda^2\!\left(\!1\!-\!\frac{A_c^2}{4}\!\right)},\nonumber\end{align}
where $\gamma_f=(\Omega-\Omega_c)/\Omega_c$ with the critical frequency $\Omega_c=\sqrt{\Omega_\text{o}^2(1-3A_c^2/4)/(1+5A_c^2/4)}$. The above result takes into account that $|\gamma_f|\ll1$.

\subsection{Switch and spin-flop fields}

The switching and spin-flop fields that are determined by measuring the EA hysteresis of the sample are used to calculate the approximate thicknesses and fringing-fields. The derivation of this formula is quite straightforward. The switching and spin-flop occur when the initial ground state is no longer an energy-minimum. Starting from the energy-density given by Eq. (\ref{W}), the switching and spin-flop fields can be determined by solving $h_x$ for
\begin{equation}
\frac{\partial^2W}{\partial\Phi^2}\frac{\partial^2W}{\partial\chi^2}-\left(\frac{\partial^2W}{\partial\Phi\partial\chi}\right)^2=0.\label{saddle}
\end{equation}
Here the out-of-plane terms are neglected since we are only interested in the in plane AP ground states in which case Eq. (\ref{saddle}) is the resulting equation. The two ground states are given by $\Phi=\Phi_0=\pi/2$ and $\chi=\chi_0=\pm\pi/2$. In this case the solutions of Eq. (\ref{saddle}) are:
\begin{equation}
\begin{aligned}
h_x&=\varepsilon(N_y-N_x-\gamma_x)\sin{\chi_0}-h_f\\
&\pm\sqrt{(N_y-N_x+\gamma_x+h_u-h_d\sin{\chi_0})^2-(1-\varepsilon^2)\gamma_y^2}.
\label{hx}
\end{aligned}
\end{equation}

Now since the dipole coupling is strong and the aspect ratio relatively close to 1, $N_y-N_x-\gamma_x<0$. Therefore the sign of $\varepsilon$ determines which of the states have positive or negative spin-flop or switch fields. Hence, in order to accurately give the relation for the different transition fields ($H_1^\pm,H_2^\pm$), equation (\ref{hx}) has to be slightly altered:
\begin{equation}
\begin{aligned}
&\frac{H_i^\pm}{4\pi M_S}=\mp(-1)^i|\varepsilon|(N_y\!-\!N_x\!-\!\gamma_x)-h_f\\
&\quad\pm\sqrt{\!\left(\!N_y\!-\!N_x\!+\!\gamma_x\!+\!h_u\!\pm\!(-1)^i \frac{\varepsilon}{|\varepsilon|}h_d\!\right)^{\!2\!}\!-\!(1-\varepsilon^2)\gamma_y^2}.
\label{Hi}
\vspace{10pt}
\end{aligned}
\end{equation}
\vspace{10pt}

Now to be able to determine the individual thicknesses and fringing fields, Eq. (\ref{Hi}) has to be rewritten in the fundamental parameters, such that none are dependent on thicknesses or fringing fields. This results in the following set of equations:
\begin{equation}
\begin{aligned}
\frac{H_i^\pm}{4\pi M_S}&=\mp(-1)^i\frac{|t_1\!-\!t_2|}{2b}(n_y\!-\!(1\!+\!r_x)n_x)\!-\!\frac{H_1^m\!+\!H_2^m}{8\pi M_S}\\
&\pm\Bigg\{\!\!\left[\!(t_1\!+\!t_2)\frac{(n_y\!-\!(1\!-\!r_x)n_x)}{2b}\!+\!\frac{H_u}{4\pi M_S}\right.\\
&\left.\pm(-1)^i \frac{t_1-t_2}{|t_1-t_2|}\frac{H_1^m-H_2^m}{8\pi M_S}\right]^{ 2\!}\!-\!t_1t_2\left(\frac{r_yn_y}{b}\right)^2\!\Bigg\}^{\!1/2},\\
\\
&\!\!\!\!\!|t_1-t_2| = \frac{H_1^+ - H_1^- - H_2^+ + H_2^-}{(n_y-(1+r_x)n_x)}\frac{b}{8\pi M_S},\\
&\!\!\!\!\!H_1^m+H_2^m=-(H_1^+ + H_1^- + H_2^+ + H_2^-)/2.\nonumber
\end{aligned}
\end{equation}

\end{document}